\documentclass[12pt]{article}

\usepackage{graphicx}

\newcommand{\be}{\begin{equation}}
\newcommand{\ee}{\end{equation}}
\newcommand{\ba}{\begin{eqnarray}}
\newcommand{\ea}{\end{eqnarray}}

\def\={\,=\,}

\def\vb0{{\bf b}_0}

\def\={\,=\,}

\begin{document} 
\thispagestyle{empty}
\begin{flushright}
WU B 12-17 \\
September, 9  2012 \\[20mm]
\end{flushright}

\begin{center}
{\Large\bf Hard exclusive wide-angle processes } \\
\vskip 10mm

P.\ Kroll \footnote{Email:  kroll@physik.uni-wuppertal.de}
\\[1em]
{\small {\it Fachbereich Physik, Universit\"at Wuppertal, D-42097 Wuppertal,
Germany}}\\
and 
{\small {\it Institut f\"ur Theoretische Physik, Universit\"at
    Regensburg, \\D-93040 Regensburg, Germany}}\\
\vskip 5mm

\end{center}
\vskip 5mm 
\begin{abstract}
In this talk the handbag approach to hard exclusive wide-angle
processes is reviewed and applications, as for instance two-photon
annihilations into pairs of mesons, are discussed.\\
Talk presented at Meson 2012, Cracow, 2012.
\end{abstract}

\section{Handbag factorization}
\label{intro}
\begin{figure}[t]
\begin{center}
\resizebox{0.5\hsize}{!}{
\includegraphics{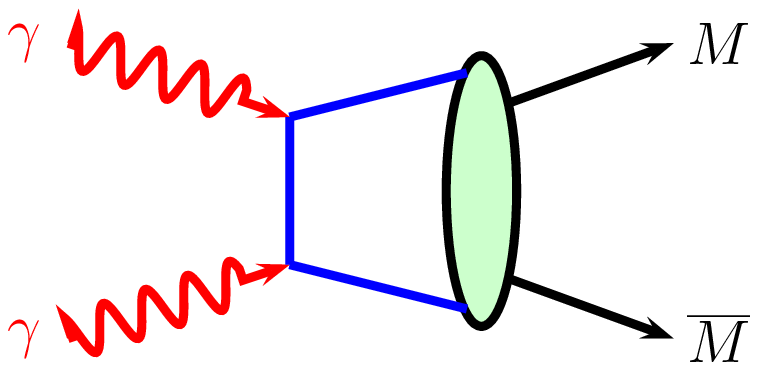}\hspace*{5em} 
\includegraphics{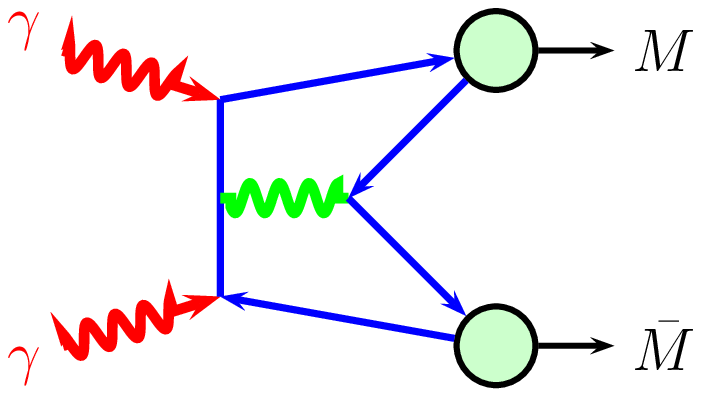}}
\end{center}
\caption{Two-photon annihilation in a meson pair within the handbag approach
(left) and within the ERBL factorization scheme (right). }
\label{fig:1}   
\end{figure}

Factorization properties of QCD allow us to calculate exclusive processes
provided a hard scale is available, either the three Mandelstam variables $s,
-t, -u$ are large as compared to a typical hadronic scale $\Lambda^2$ where
$\Lambda$ is of order $1\,{\rm GeV}$ (wide-angle processes) or there is a highly
virtual photon involved (deeply virtual processes). In the space-like region 
typical hard processes are real and virtual Compton scattering or photo- and 
electroproduction of mesons. In these cases the process amplitudes factorize 
in a hard partonic subprocess (e.g. $\gamma^{(*)}q\to \gamma q$) and in soft 
hadronic matrix elements parameterized as generalized parton distributions
(GPDs). For time-like processes, e.g.\ two-photon annihilations into pairs of 
hadrons, an analogous factorization scheme holds 
\cite{pire,DKV2,DKV3,weiss,DK}. In this case the soft hadronic matrix elements 
are so-called two-hadron distribution amplitudes, time-like versions of GPDs, 
see Fig.\ \ref{fig:1}. 

One may consider more complicated topologies than shown in Fig.\ \ref{fig:1} 
in which $n$ partons are emitted
and reabsorbed from the hadrons. In order to have quasi-on-shell partons
entering the hadrons, a compelling requirement for factorization, $n-1$
hard gluons are needed to be exchanged between the $n$ active partons. For
very large $-t$ and $-u$ one may treat the hadrons in valence quark
approximation. Considering for instance $\gamma\gamma\to M\overline{M}$ with an
active quark-antiquark pair ($n=2$). In this case there is no spectator left 
and, hence, the soft physics is encoded in two meson distribution amplitudes
instead of a  time-like GPD. This is an example of the so-called ERBL 
factorization scheme which has been invented for $\gamma\gamma\to M\bar M$ in \cite{BL89}.

The arguments for factorization of $\gamma\gamma\to M\overline{M}$ in the 
wide-angle region have been given in \cite{DKV2}. For the ease of
legibility let $M$ be charged pion; the generalization to other pseudoscalar
mesons is straightforward. It is of advantage to work in a symmetric (c.m.)
frame in which the pions move along the 1-axis. The light-cone plus components
of the meson momenta, $p$ and $p^\prime$, are then equal: $p^+=p^{\prime +}$
and the skewness, defined by $\zeta=p^+/(p^++p^{\prime +})$, is 1/2. The momentum
fraction of the active quark is as usual defined by 
$z=k^+/(p^++p^{\prime +})$; the momenta fraction of the antiquark is
$\bar{z}=1-z$. In order to achieve the factorization in a hard process 
$\gamma\gamma\to q\bar q$ and a soft transition $q\bar q\to \pi^+\pi^-$, two 
assumptions are made in \cite{DKV2}: i) restricted transverse momenta,
$k_{\perp i}/z_i \sim \Lambda^2$, for the active partons as well for the
spectators, and ii) all virtualities at the parton-hadron vertices are soft of
order $\Lambda^2$. With the help of these assumptions one can show that the
following requirements must hold: 
\begin{equation}
2z-1,\quad \sin{\varphi} \sim \Lambda^2/s
\label{eq:requirements}
\end{equation}
where $\varphi$ describes the orientation of the active parton momenta in the
1--2 plane. The second requirement has two solutions 
\begin{equation}
\varphi \simeq 0: \quad k\simeq p\,, \quad k^\prime\simeq p^\prime \qquad
\varphi \simeq \pi: \quad k\simeq p^\prime\,, \quad k^\prime\simeq p 
\end{equation}
One can then derive a factorization formula for the helicity amplitudes
\begin{equation}
{\mathcal A}_{\mu\mu'}= -\sum_q(ee_q)^2 \int \frac{d^4k}{\sqrt{k^+k^{'+}}}\, {\mathcal
  H}_{\mu\mu'}(k,k')\, S(k,k') + \mbox{axial current term}
\label{eq:ansatz}
\end{equation}
where due to charge conjugation invariance
\begin{equation}
S(k,k')=-S(k',k)\,, \qquad {\mathcal H}(k,k')=-{\mathcal H}(k',k)\,.
\end{equation}
The soft matrix element $S(k,k')$ is expected to be strongly peaked when   
(\ref{eq:requirements}) is fulfilled. The two regions $k\sim p$ and $k\sim p'$
where this is the case are related through a rotation by $\pi$ about the
3-axis of our coordinate frame. The hard scattering kernel ${\mathcal H}$ can 
be Taylor expanded around the 1-axis and $z=1/2$ 
\begin{equation}
{\cal H}_{\pm\mp} = 2\Big(\sqrt{u/t}-\sqrt{t/u}\Big)\, - \,(z-\bar z) (s/t+u/t)\, 
+ \, {\cal O}\Big((z-\bar z)^2,\varphi^2\Big)\,,
\label{eq:exp}
\end{equation}
where the labels denote the helicities of the photons (${\cal H_{\pm\pm}}=0$). 
In terms of the c.m.\ scattering angle, $\theta$, the first term in (\ref{eq:exp}) is
proportional to $1/\sin{\theta}$ while the second one is $\propto 1/\sin^2{\theta}$. 
It can then be shown \cite{DKV2} that for the $\pi^+\pi^-$ channel the first
term in this expansion vanishes due to a conspiracy of charge conjugation and
rotation by $\varphi=\pi$. Keeping  therefore only the second term in
(\ref{eq:exp}) and perform the integrals in (\ref{eq:ansatz}) one arrives at
\begin{equation}
{\mathcal A}_{+-}={\mathcal A}_{-+}= -4\pi \alpha_{\rm elm} \frac{s^2}{tu}
R_{2\pi}(s)\,,
\end{equation}
where the annihilation form factors encoding the soft physics, is defined by
\begin{equation}
R_{2\pi}(s) = \sum_{q=u,d,s} e_q^2\, R_{2\pi}^q(s)\,, \qquad 
R^q_{2\pi}(s)=\frac12\int_0^1 dz\,(2z-1)\,\Phi_{2\pi}^q(z,\zeta=1/2,s)\,.
\label{eq:annFF}
\end{equation}
Here, $\Phi^q_{2\pi}$ is the two-pion distribution amplitude \cite{pire} defined as the
Fourier transform of a bilocal vacuum-two-pion matrix element of quark field operators. 
This distribution amplitude also determines the electromagnetic form
factor of the pion in the time-like region
\begin{equation}
F_\pi(s) =e_u F^u_\pi(s) + e_d F^d_\pi(s)\,, \qquad F_\pi^q(s)=\int_0^1
dz\,\Phi^q_{2\pi}(z,\zeta=1/2,s)\,.
\end{equation}
The energy dependences of the annihilation and electromagnetic form factors
are not predicted in the handbag approach.
 
Although the $z-\bar{z}$ term in (\ref{eq:exp}) is of the same order as the
parton-off-shell effects we remain with the on-shell approximation. Therefore the
above sketched approach to $\gamma\gamma\to\pi\pi$ is to be considered as a
model. The suppression of the leading term in (\ref{eq:exp}) is special to
$\gamma\gamma\to \pi\pi$; it is due to a conspiracy of charge conjugation
invariance and a rotation. This conspiracy does not occur in other reactions
and the leading term in (\ref{eq:exp}) dominates. For instance, in two-photon
annihilations into baryon-antibaryon ($B\bar{B}$) pairs \cite{DKV3} the region 
$k'\simeq p$ corresponds to antiquark hadronization into a baryon which
requires sea quarks with very high momentum fractions. It is unlikely that
such sea quarks exist. Another example is set by real Compton scattering in
the space-like region \cite{DFJK1}. The regions $k^+>0$ ($\gamma q\to\gamma
q$) and $k^+ <0$ ($\gamma \bar{q}\to \gamma \bar{q}$) are related by charge
conjugation but not by a rotation.

The above handbag result for the $\pi\pi$ channel can easily be generalized to
the production of other pairs of pseudoscalar mesons. The corresponding
differential cross section reads
\begin{equation}
\frac{d\sigma}{dt}(\gamma\gamma\to M\overline{M})=\frac{8\pi\alpha_{\rm elm}}{s^2}
                 \,\frac1{\sin^4{\theta}}\,\left|R_{M\overline{M}}(s)\right|^2\,.
\end{equation}
There are six
pseudoscalar meson channels available in two-photon annihilations including
altogether 18 annihilation form factors. To fix these form factors from
experiment is a boring program. However, due to flavor symmetry (the meson
pair couples to an $U$-spin singlet) and due to the
absence of isospin 2 (and $V$-spin 2) states (because the two photons annihilate 
via a quark-antiquark intermediate state) there are only two independent
annihilation form factors, say $R_{2\pi}^u$ and $R_{2\pi}^s$, or, in other
words, a valence and a non-valence form factor. The combinations of the
individual flavor contributions appearing in the various channels read in
terms of the two independent form factors
\begin{eqnarray}
R_{\pi^+\pi^-}&=&R_{\pi^0\pi^0}=R_{K^+K^-} = \frac59 R_{2\pi}^u + \frac19 R_{2\pi}^s\,, \nonumber \\
R_{K^0\overline{K}{}^0}&=&\frac29 R_{2\pi}^u +  \frac49 R_{2\pi}^s\,,\qquad
R_{\eta\pi^0}=\frac1{3\sqrt{3}}( R_{2\pi}^u - R_{2\pi}^s)\,,  \nonumber\\  
R_{\eta\eta}&=&\frac13( R_{2\pi}^u +  R_{2\pi}^s)\,.
\label{eq:symmetry_relations}
\end{eqnarray}

In addition there are a number of relations and triangular inequalities
among the cross sections. 
\section{Results on $\gamma\gamma\to M\overline{M}$}
\begin{figure}[t]
\begin{center}
\resizebox{0.34\hsize}{!}{
\includegraphics{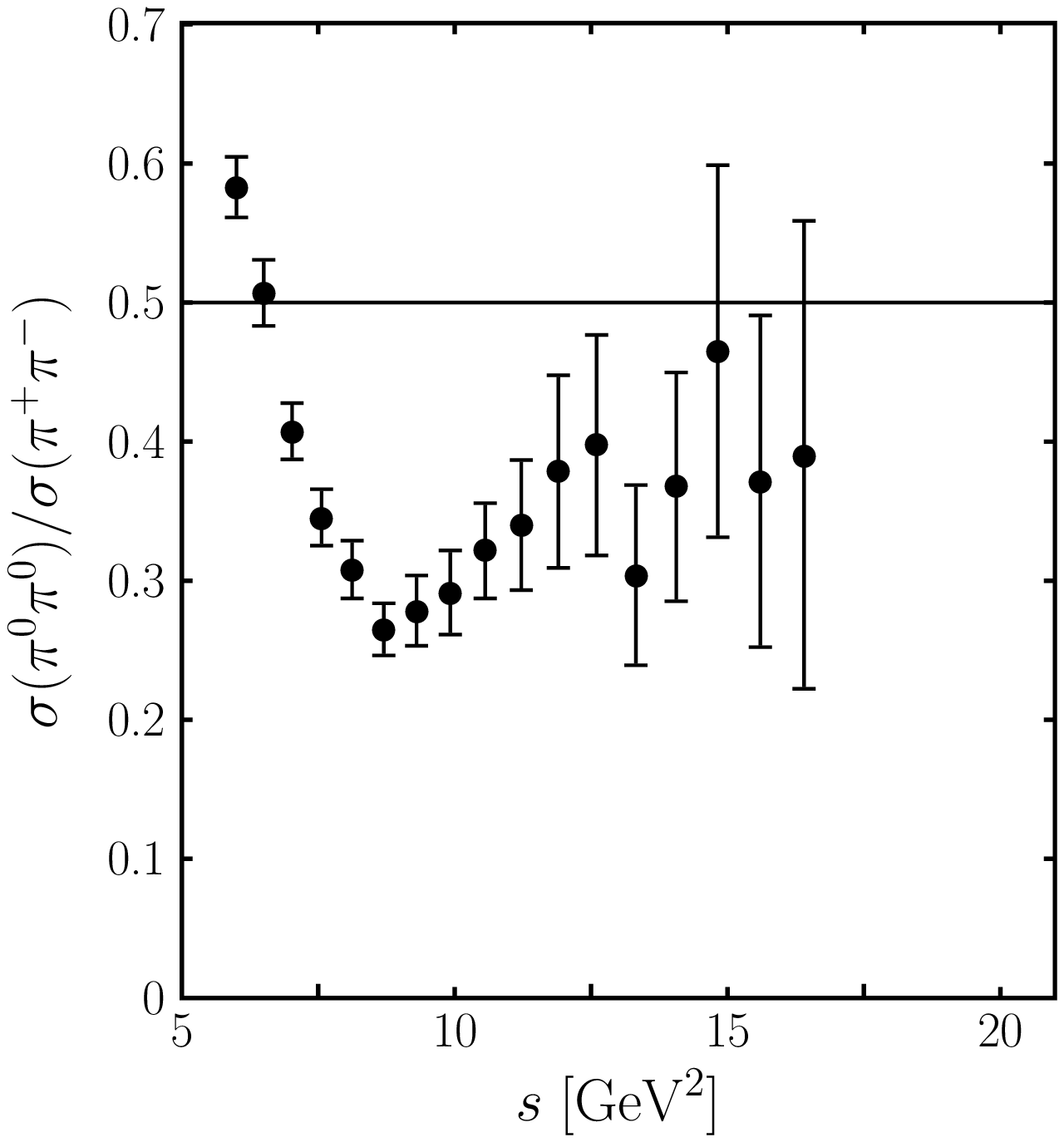}}\hspace*{1em} 
\resizebox{0.28\hsize}{!}{
\includegraphics{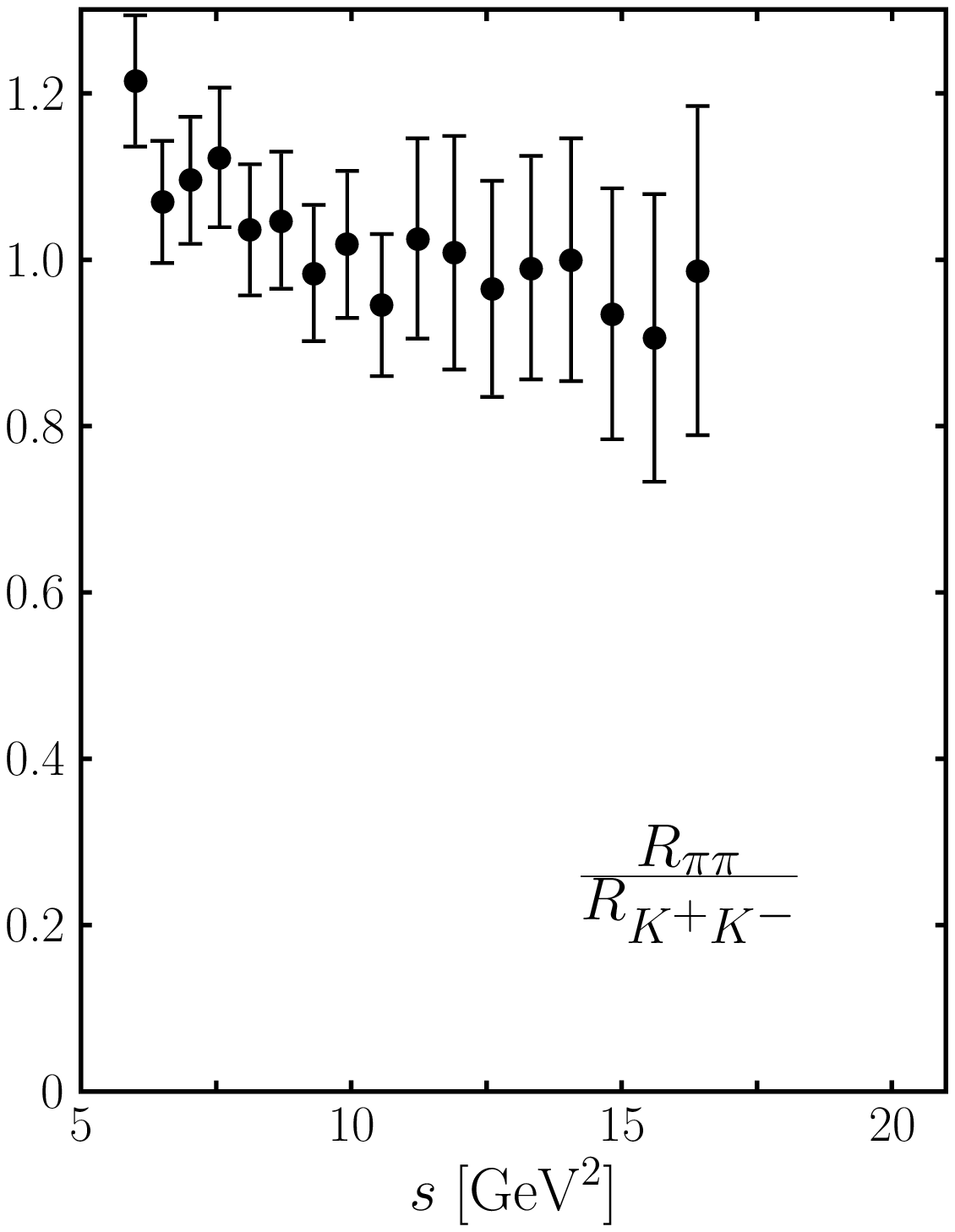}}\hspace*{1em}
\resizebox{0.31\hsize}{!}{
\includegraphics{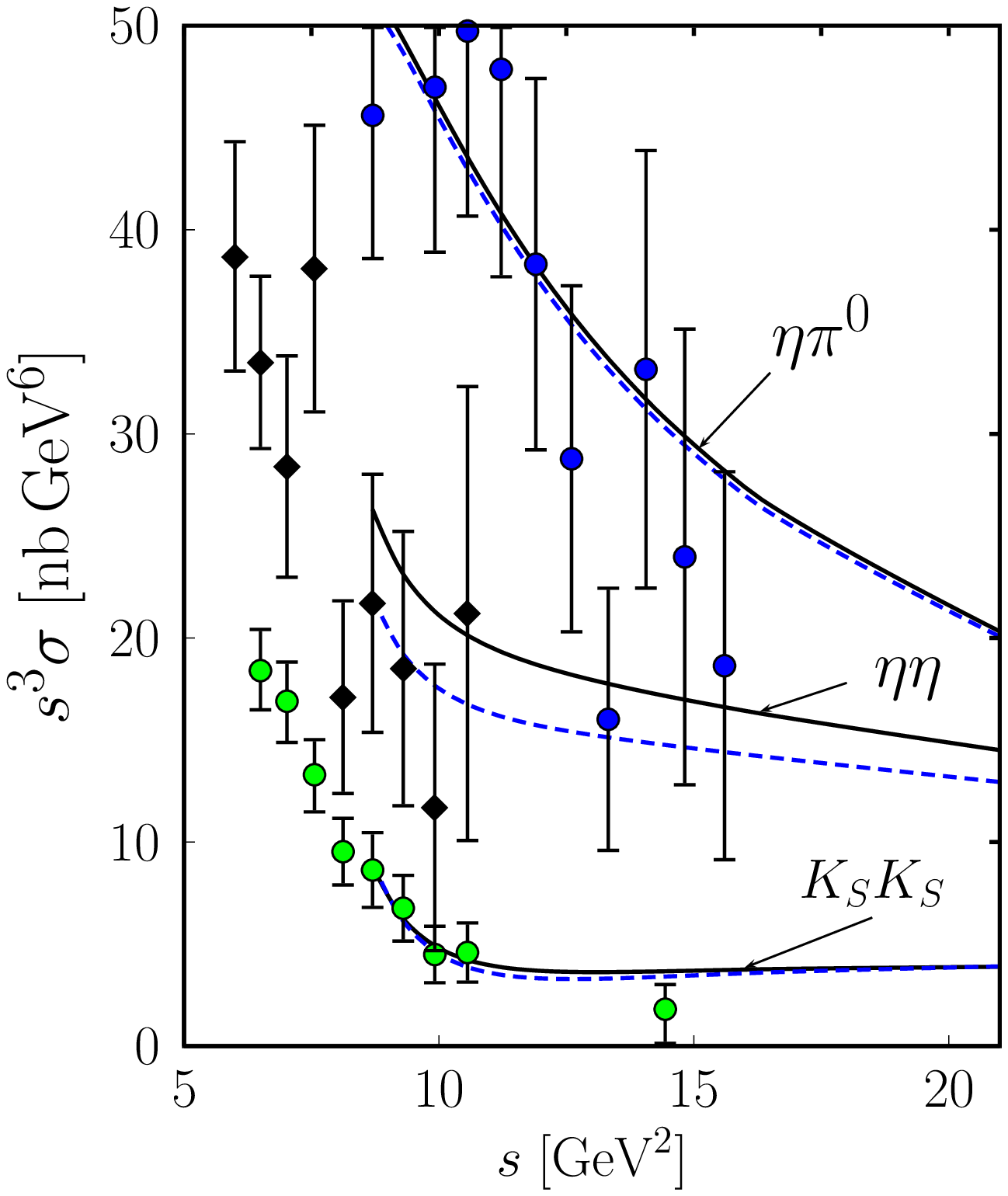}} 
\end{center}
\caption{Left: Cross section ratio of neutral and charged pion pair
  production. Data taken from \cite{BELLE-pi0}. Mid: Ratio of $\pi\pi$ and
  $K^+K^-$ annihilation form factors. Data taken from \cite{BELLE-piK}. Right:
  Data \cite{BELLE-KS,BELLE-etapi0,BELLE-etaeta} and predictions for various
  cross sections integrated over $|\cos{\theta}|\leq 0.6$. Dashed lines: 
  fit with $\eta-\eta^\prime$ mixing is taken into account.}
\label{fig:ratios}   
\end{figure}

The BELLE collaboration has measured the wide-angle cross sections for the six
pseudoscalar meson channels up to fairly large energies
\cite{BELLE-piK,BELLE-pi0,BELLE-KS,BELLE-etapi0,BELLE-etaeta}. The data are
compatible  with a $1/\sin^4{\theta}$ behavior for $s> 9\,\mbox{GeV}^2$.
Given that fact one has only to deal with the form factors or the respective
integrated cross sections. It turns out that the inequalities and relations
following from flavor symmetry and isospin selection rules are satisfied
within errors in general. From (\ref{eq:symmetry_relations}) and from 
the statistical factor to be applied to the integrated cross section for identical
particles it follows that the ratio of the $\pi^0\pi^0$ and $\pi^+\pi^-$
is predicted to be $1/2$ in the handbag approach. The data shown in Fig.\
\ref{fig:ratios} clearly deviate from $1/2$ at low $s$ but seem to approach
this value for $s\geq 9\,\mbox{GeV}^2$ within rapidly growing errors. A small
$I=2$ admixture which may vanish for $s\to\infty$, can easily explain the
deviation from $1/2$. Assuming for example that the $I=2$ and $I=0$ amplitudes 
are in phase then an admixture ${\mathcal A}^{I=2}<0.11 {\mathcal A}^{I=0}$ is
sufficient to explain the observed deviation. Thus, the dominance of the
$I=0$ amplitude is fully consistent with experiment for large $s$. In Fig.\ 
\ref{fig:ratios} the ratio of the $\pi\pi$ combination
$1/3(\sigma(\gamma\gamma\to\pi^0\pi^0)+\sigma(\gamma\gamma\to\pi^+\pi^-))$
and the $K^+K^-$ cross sections are shown and can be seen to be compatible
with 1 according to the handbag approach, see (\ref{eq:symmetry_relations}). 
Note that in the above combination of $\pi\pi$ cross sections the interference 
term between the $I=2$ and $I=0$ cancels.

For a few values of the energy the two basic form factors as well as their
relative phase, $\rho$, can directly be extracted from the data on, say, $K^+K^-$,
$K^0\bar{K}{}^0$ and $\eta\pi^0$ production. Alternatively, in order to reduce
errors, one may perform a combined energy-dependent fit. This fit provides
\begin{eqnarray}
s|R^u_{2\pi}| &=& 1.37\,\mbox{GeV}^2\,\left(\frac{s_0}{s}\right)^{0.42}\,,
\quad
s|R^s_{2\pi}| = 0.50\,\mbox{GeV}^2\,\left(\frac{s_0}{s}\right)^{1.22}\,, \nonumber\\
\rho &=& \pi \Big[1+tanh{\big(\frac{0.63\,\mbox{GeV}^2}{s-6.0\,\mbox{GeV}^2}\big)}\Big]
\label{eq:parameter}
\end{eqnarray}
($s_0=9\,\mbox{GeV}^2$). The energy dependencies of the other channels
($\pi^+\pi^-$, $\pi^0\pi^0$, $\eta\eta$) are then fixed in agreement with
experiment. The relative phase varies rapidly in the charmonium region 
and approaches $\pi$ for $s\to\infty$. In concord with expectation, the 
non-valence form factor, $R^s_{2\pi}$, is much smaller than the valence one 
and decreases more rapidly with increasing energy than the other one. Hence, 
for $s\to\infty$ only the valence form factor remains and the six cross 
sections all exhibit the same energy dependence and only differ in magnitude 
by charge factors. For the range of energy measured by BELLE the energy 
dependence of the various cross sections differ quite a bit due to the
different superpositions of the valence and non-valence form factors.
Charged pions and Kaons have the smallest contributions from the non-valence 
form factor. Therefore, their integrated cross sections drop only slightly 
faster then $s^{-3}$. On the other hand, the $K^0\bar{K}{}^0$ ($ = 2K_SK_S$) 
cross section falls off rapidly with energy because of the relative strong
contribution from the non-valence form factor in this case. This is in 
agreement with experiment as can be seen from Tab.\ \ref{tab:1} where the 
effective powers of $s$ obtained by BELLE from fits 
$\sigma\propto s^{-n_{\rm eff}}$ to its data are shown.

\begin{table*}[bh]
\renewcommand{\arraystretch}{1.4} 
\begin{center}
\begin{tabular}{|c|c | c ||c | c | c |}
\hline 
 channel  &  $n_{\rm eff}=$ & $|\cos{\theta}|<$ & channel&  $n_{\rm eff}=$ 
        & $|\cos{\theta}|<$  \\[0.2em]\hline 
 $\pi^+\pi^-$  & $4.0(0.2)(0.7)$ & 0.6 & $\eta\eta$ & $3.9 (0.3)(0.2)$  & 0.8 \\[0.2em]
 $K^+K^-$  & $3.7(0.2)(0.7)$ & 0.6 &$\eta\pi^0$ & $5.3(0.6)(0.3)$ & 0.8 \\[0.2em]
$\pi^0\pi^0$ & $4.0(0.2)(0.7)$ & 0.8 & $K_SK_S$   & $5.3(0.3)(0.2)$   & 0.6 \\[0.2em]  
\hline
\end{tabular}
\end{center}
\caption{Effective powers of the integrated cross sections measured by BELLE
  \cite{BELLE-piK}--\cite{BELLE-etaeta}.}
\label{tab:1}
\renewcommand{\arraystretch}{1.0}   
\end{table*} 
 
The process of interest has also been investigated in the ERBL factorization
scheme \cite{BL89,chernyak}. In this approach the mesons are treated in
valence quark approximation and the amplitudes are given by a
convolution of the distribution amplitudes for the mesons and a hard
scattering kernel
\begin{equation}
{\mathcal A} \sim f_{M_1} f_{M2} \int dx\, \Phi_{M_1}(x) \int dy\, \Phi_{M_2}(y)
T(x,y,s,\theta)
\end{equation}
where $f_{M_i}$ is the decay constant of meson $M_i$. As for the handbag approach
the cross section behaves as $1/\sin^4{\theta}$ but its energy dependence is
predicted to be $\sigma \propto s^{-3}$, cf.\ Tab.\ \ref{tab:1}. While this is
in rough agreement with experiment for charged mesons the magnitude of the 
corresponding cross section is underestimated when it is evaluated from a 
distribution amplitude that is close to the asymptotic form, $6x(1-x)$ which
is favored by current phenomenology and lattice results \cite{braun}. The
production of neutral meson pairs is generically suppressed, since at leading
order of $\alpha_s$ the bulk of the amplitude is proportional to a charge factor 
$(e_{q_1}-e_{q_2})^2$ for a meson with quark content $q_1\bar{q}_2$ \cite{BL89}. 
Explicit calculations \cite{BL89,chernyak} yield values below 0.05 for the 
ratio of $\pi^0\pi^0$ and $\pi^+\pi^-$  cross sections which is significantly 
below the experimental results shown in Fig.\ \ref{fig:ratios}. Because of
charge factors the ratio of $K_SK_S$ and $K^+K^-$ cross sections is predicted 
to be even smaller. Despite this suppression of the neutral meson pairs the
energy dependencies of the corresponding cross sections is $\propto s^{-3}$
which is also in conflict with experiment, see Tab.\ \ref{tab:1}. For these
reasons it was suggested in \cite{chernyak} that at BELLE energies the
production of neutral meson pairs is dominated by a contribution other then the
the ERBL mechanism but a quantitative study of the new contribution is lacking.
\section{Other two-photon processes}
\begin{figure}[t]
\begin{center}
\resizebox{0.25\hsize}{!}{
\includegraphics{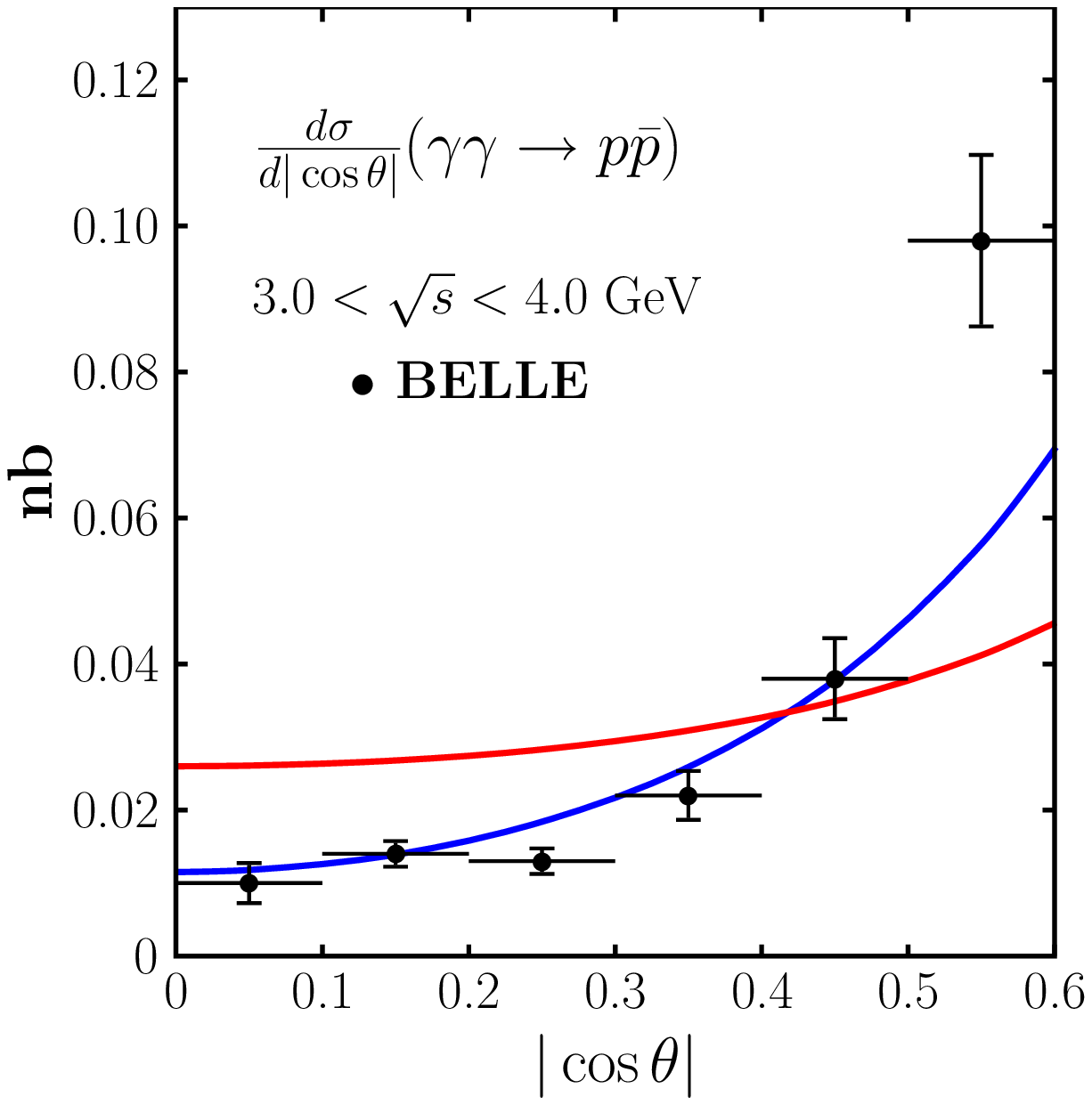}}\hspace*{1em}
\resizebox{0.32\hsize}{!}{
\includegraphics{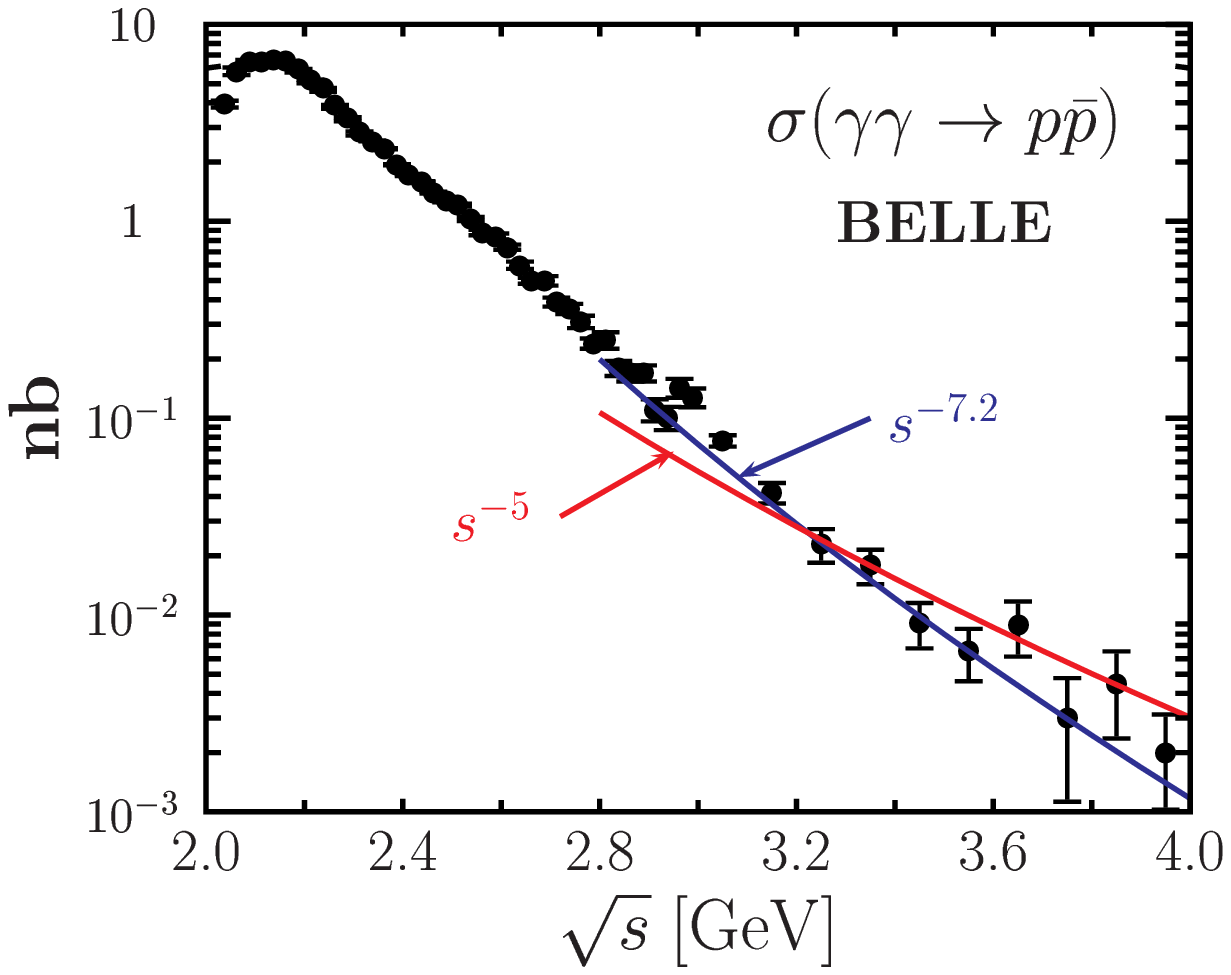}} \hspace*{1em}
\resizebox{0.32\hsize}{!}{
\includegraphics{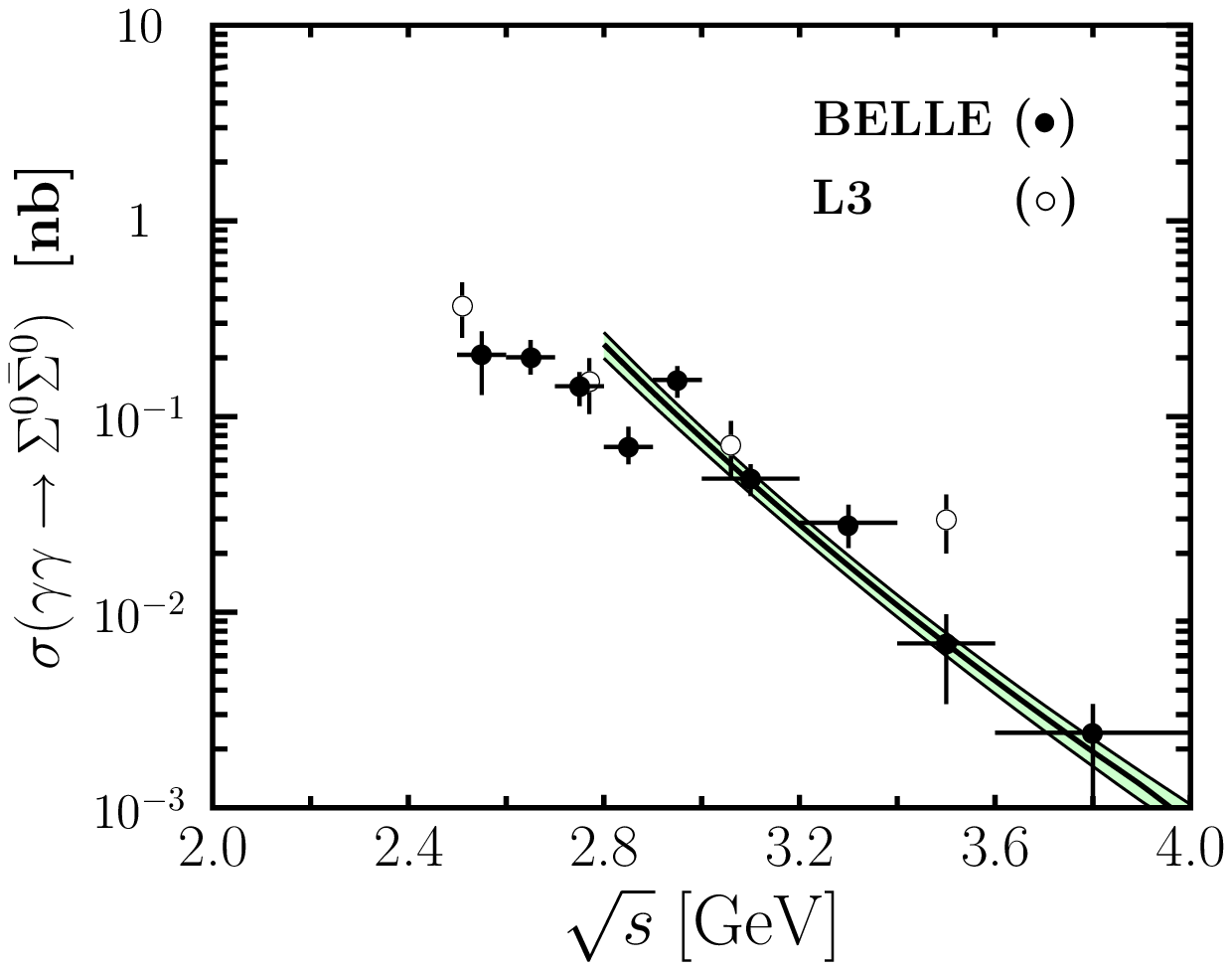}}
\end{center}
\caption{The differential (left) and integrated (mid) cross sections for $p\bar{p}$ 
production. Data taken from \cite{BELLE-pbarp}. Blue solid lines
  represent the handbag results \cite{schaefer}. Right: Integrated cross section
  for $\Sigma^0\overline{\Sigma}^0$. The data from \cite{BELLE-LL-SS,L3} are compared to
  the handbag result.} 
\label{fig:pp}   
\end{figure}

The meson case can be straightforwardly generalized to the production
of baryon-antibaryon pairs. An important difference is
that the cross section behaves $\propto 1/\sin^2{\theta}$ as has been discussed
at the of Sect.\ 1 (see also (\ref{eq:exp})). Due to the baryon spin there are
four GPDs and hence, four form factors of which one decouples in the symmetric
frame. Thus, the differential cross section for the production of a
$B\overline{B}$ reads \cite{DKV3}
\begin{equation}
\frac{d\sigma}{dt}(\gamma\gamma\to B\overline{B})=\frac{4\pi\alpha_{\rm elm}}
                  {s^6\sin^2{\theta}}\,\left\{\big|s^2R^{\,\rm eff}_{B\overline{B}}\big|^2
                  + \cos^2{\theta}\big|s^2R^{\,V}_{B\overline{B}}\big|^2\right\}\,,
\end{equation}
where
\begin{equation}
R^{\,\rm eff}_{B\overline{B}} =\sqrt{\big|R^{\,A}_{B\overline{B}}+R^{\,P}_{B\overline{B}}\big|^2
               +\frac{s}{4m^2}\big|R^{\,P}_{B\overline{B}}\big|^2} 
\end{equation}
The axial and the pseudoscalar form factors can be disentangled with
polarization measurements. As for the meson case (see (\ref{eq:annFF}))  each
form factor is a sum of individual flavor form factors ($i=V, A, P$)
\begin{equation}
R^{\,i}_{B\overline{B}} = \sum_{q=u,d,s} e_q^2 F^{\,iq}_{B\overline{B}}\,, \qquad
F^{\,iq}_{B\overline{B}}(s)=\int_0^1 dz \Phi^{\,iq}_{B\overline{B}}(z,1/2,s)\,,
\end{equation}
where $\Phi^{\,iq}_{B\overline{B}}$ is a baryon-antibaryon distribution amplitude.
For comparison the magnetic form factor of the proton reads
\begin{equation}
G_M^{\,p} = \sum_{q=u,d,s} e_q^2 F^{\,Vq}_{p\bar{p}}\,,
\end{equation}
In Fig.\ \ref{fig:pp} the BELLE data \cite{BELLE-pbarp} on $\gamma\gamma\to
p\bar{p}$ are compared to the handbag results \cite{schaefer}. The 
effective and the vector form factors are parameterized analogously to
(\ref{eq:parameter}) and the parameters fitted to the data. From Fig.\ \ref{fig:pp} 
one sees that the data are nicely compatible with a $1/\sin^2{\theta}$
behavior. The energy dependence of the form factors is $s^{-3.1}$, i.e.\ the
integrated cross section falls off as $s^{-7.2}$ at large $s$. Note that in
the ERBL approach the cross section behaves $\propto s^{-5}$. 
  
As for the case of mesons one can show that due to the absence of $I=V=2$
states and $U$-spin symmetry there are only three independent flavor form 
factors for the ground state baryon channels for which one may take the ones
for the proton-antiproton channel, $F^{\,iq}_{p\bar{p}}$. The quality of the
present data \cite{BELLE-LL-SS,L3} necessitates a simplifying assumption
\begin{equation}
\rho_d = F^{\,id}_{p\bar{p}}/F^{\,iu}_{p\bar{p}}\,, \qquad  
              \rho_s = F^{\,is}_{p\bar{p}}/F^{\,iu}_{p\bar{p}}\,,  
\end{equation}
which leads to 
\begin{equation}
\sigma(\gamma\gamma\to B\overline{B}) = r_B^2\,\sigma(\gamma\gamma\to
p\bar{p})\,,
\end{equation}
with
\begin{equation}
r_\Lambda=-\frac32\,\frac{1+2\rho_d+\rho_s}{4+\rho_d+\rho_s}\,, \qquad
r_{\Sigma^0}=-\frac12\,\frac{5+2\rho_d+5\rho_s}{4+\rho_d+\rho_s}\,.
\end{equation}
A fit to the data \cite{BELLE-LL-SS,L3} provides the results shown in Fig.\
\ref{fig:pp} ($\rho_d=0.75$, $\rho_s=-4.1\,\mbox{GeV}^2/s$). Similar results
are obtained for the $\Lambda\overline{\Lambda}$ channel.

Due to time-reversal invariance the amplitudes for $\gamma\gamma\to p\bar{p}$
and $p\bar{p}\to\gamma\gamma$ are the same up to a phase. This offers the
opportunity for the FAIR project to study also handbag physics in the
wide-angle region. In addition to $p\bar{p}\to\gamma\gamma$ one may study the 
photon-meson channels. Their amplitudes are still under control of the basic 
form factors $F^{\,iq}_{p\bar{p}}$. Thus, for instance, for the $\gamma\pi^0$
channel the annihilation form factors read  \cite{schaefer} ($i=V,A,P$)
\begin{equation}
R^{\,i}_{\pi^0}=\frac1{e_u\sqrt{2}}\,
   \frac{1-e_d/e_u\rho_d}{1+(e_d/e_u)^2\rho_d+(e_s/e_u)^2\rho_s}\,R^{\,i}_{p\bar{p}} \,.
\end{equation}
With these form factors one finds good agreement with the
FermiLab data \cite{E760} on $p\bar{p}\to \gamma\pi^0$.

Finally, it should mentioned that the handbag approach to wide-angle
scattering also applies to space-like processes. The gold-plated example is 
real Compton scattering (cf.\ the remarks in Sect.\ \ref{intro}) for which the
GPDs or respective form factors are known from an analysis of the nucleon form
factors \cite{DFJK4}. The predictions for real Compton scattering are in
remarkable agreement with experiment.

\section{Summary}
The basic ideas of the handbag approach for wide-angle scattering are reviewed
and applications to several time-like processes such as two-photon
annihilations into pairs of mesons or baryon-antibaryons, are discussed. It
turns out that this approach seems to work quite well for energies larger than
about 3 GeV. Data at higher energies are highly welcome in order to have a
clean suppression of subleading terms. In the near future 
proton-antiproton annihilations into two photons or photon-meson can be
measured at FAIR. Such data in combination with $\gamma\gamma\to
B\overline{B}$ measurements performed at BELLE II would allow for a 
detailed investigation of the proton-antiproton distribution amplitudes.

{\it\bf Acknowledgement} The author wishes to thank the organizers of the
MESON 2012 workshop for inviting him to present this talk.

\end{document}